\documentclass[aps,prxquantum,reprint,twocolumn,superscriptaddress,floatfix,nofootinbib,longbibliography]{revtex4-1}
\usepackage{graphicx,amsmath,amsfonts,amssymb,amsthm,xr}
\usepackage{epsfig,amsmath,amssymb,color,dsfont,upgreek,physics}
\usepackage{mathrsfs}
\usepackage{mathtools}
\usepackage{bbold}
\usepackage{comment}
\usepackage{float}

\usepackage[bookmarks=true,colorlinks,linkcolor=OrangeRed,urlcolor=NavyBlue,citecolor=RoyalBlue]{hyperref}
\usepackage[dvipsnames]{xcolor}
\usepackage{orcidlink}

\definecolor{mygold}{rgb}{0.93,0.69,0.13}

\definecolor{mypurple}{rgb}{0.49,0.18,0.56}

\definecolor{mygreen}{rgb}{0,0.5,0}

\definecolor{mygreen}{rgb}{0,0.5,0}
\definecolor{myred}{rgb}{0.7,0,0}
\definecolor{myblue}{rgb}{0,0,1}

\begin{document}
\title{Spin-$S$ $\mathrm{U}(1)$ Quantum Link Models with Dynamical Matter on a Quantum Simulator}
\author{Jesse Osborne${}^{\orcidlink{0000-0003-0415-0690}}$}
\email{j.osborne@uqconnect.edu.au}
\affiliation{School of Mathematics and Physics, The University of Queensland, St.~Lucia, QLD 4072, Australia}
\author{Bing Yang${}^{\orcidlink{0000-0002-8379-9289}}$}
\email{yangbing@sustech.edu.cn}
\affiliation{Department of Physics, Southern University of Science and Technology, Shenzhen 518055, China}
\author{Ian P.~McCulloch${}^{\orcidlink{0000-0002-8983-6327}}$}
\affiliation{School of Mathematics and Physics, The University of Queensland, St.~Lucia, QLD 4072, Australia}
\author{Philipp Hauke${}^{\orcidlink{0000-0002-0414-1754}}$}
\email{philipp.hauke@unitn.it}
\affiliation{Pitaevskii BEC Center and Department of Physics, University of Trento, Via Sommarive 14, I-38123 Trento, Italy}
\affiliation{INFN-TIFPA, Trento Institute for Fundamental Physics and Applications, Trento, Italy}
\author{Jad C.~Halimeh${}^{\orcidlink{0000-0002-0659-7990}}$}
\email{jad.halimeh@physik.lmu.de}
\affiliation{Department of Physics and Arnold Sommerfeld Center for Theoretical Physics (ASC), Ludwig-Maximilians-Universit\"at M\"unchen, Theresienstra\ss e 37, D-80333 M\"unchen, Germany}
\affiliation{Munich Center for Quantum Science and Technology (MCQST), Schellingstra\ss e 4, D-80799 M\"unchen, Germany}

\begin{abstract}
Quantum link models (QLMs) offer the realistic prospect for the practical implementation of lattice quantum electrodynamics (QED) on modern quantum simulators, and they provide a venue for exploring various nonergodic phenomena relevant to quantum many-body physics. In these models, gauge and electric fields are represented by spin-$S$ operators. So far, large-scale realizations of QLMs have been restricted to $S=1/2$ representations, whereas the lattice-QED limit is approached at $S\to\infty$. Here, we present a bosonic mapping for the representation of gauge and electric fields with effective spin-$S$ operators for arbitrarily large values of $S$. Based on this mapping, we then propose an experimental scheme for the realization of a large-scale spin-$1$ $\mathrm{U}(1)$ QLM using spinless bosons in an optical superlattice. Using perturbation theory and infinite matrix product state calculations, which work directly in the thermodynamic limit, we demonstrate the faithfulness of the mapping and stability of gauge invariance throughout all accessible evolution times. We further demonstrate the potential of our proposed quantum simulator to address relevant high-energy physics by probing the (de)confinement of an electron--positron pair by tuning the gauge coupling. Our work provides an essential step towards gauge-theory quantum simulators in the quantum-field-theory limit.
\end{abstract}

\date{\today} 
\maketitle
\tableofcontents

\section{Introduction}
In recent years, the quantum simulation of gauge theories has emerged as an exciting research field in quantum many-body physics \cite{Pasquans_review,Dalmonte_review,Zohar_review,aidelsburger2021cold,Zohar_NewReview,Bauer_review,funcke2023review}. Prominent in the context of high-energy physics, gauge theories describe how elementary particles interact through the mediation of gauge bosons, and they form the bedrock of the Standard Model \cite{Weinberg_book,Gattringer_book,Cheng_book}. They are also of great relevance in condensed matter physics. For example, they can be used to describe strongly correlated quantum matter with fractionalized excitations such as spin liquids \cite{Balents_NatureReview,Savary2016}.

As with most quantum many-body systems, gauge theories are difficult to simulate using classical methods. Indeed, numerical investigations of gauge theories are restricted either to exact results for very small system sizes or short evolution times for large sizes using matrix product state (MPS) techniques \cite{Dalmonte_review,Uli_review}. Quantum Monte Carlo simulations suffer from the sign problem, which limits their ability in calculating real-time dynamics.
Following Richard Feynman's vision \cite{Feynman1982}, a way out may be found by developing quantum simulators to model quantum many-body systems, and in recent years there has been impressive progress in this endeavor \cite{Bloch2008,Trabesinger2012,Hauke2012,Georgescu_review}. In particular, gauge-theory quantum simulators have witnessed many experimental realizations on various platforms \cite{Martinez2016,Muschik2017,Bernien2017,Klco2018,Kokail2019,Schweizer2019,Goerg2019,Mil2020,Klco2020,Yang2020,Atas2021,Zhou2022,Nguyen2021,Wang2021,Mildenberger2022,Wang2022,charles2023simulating}, opening the door to the possibility of probing high-energy physics on table-top quantum devices.

A prominent gauge theory of great theoretical and experimental relevance is quantum electrodynamics (QED). For the purpose of numerical and experimental feasibility, a \textit{quantum link} formulation of QED has been proposed, where the gauge and electric-field operators are represented by spin-$S$ operators \cite{Chandrasekharan1997}. The resulting spin-$S$ $\mathrm{U}(1)$ quantum link models (QLMs) retrieve quantum electrodynamics (QED) in the limit of $S\to\infty$ \cite{Chandrasekharan1997,Wiese_review}. Recently, the method of linear gauge protection \cite{Halimeh2020e,Lang2022SGP} has facilitated the stabilization of gauge invariance in quantum simulations of the spin-$1/2$ $\mathrm{U}(1)$ QLM, leading to corresponding experimental realizations in large-scale Bose--Hubbard quantum simulators \cite{Yang2020,Zhou2022,Wang2022}. However, in order to approach QED, larger values of $S$ are required experimentally. This motivates the development of experimentally feasible proposals for the implementation of $\mathrm{U}(1)$ QLM at $S>1/2$. 
One way is based on small Bose--Einstein condensates representing link variables \cite{Kasper2017} in conjunction with imposing gauge-invariance through angular momentum conservation \cite{Zohar2013PRA,Stannigel2014}. Such a scheme has been demonstrated experimentally for a single building block in the regime of large $S$ \cite{Mil2020}. 
Alternative approaches are based on using highly occupied bosonic modes \cite{Zohar2011,Yang:2017pra,Ott2020scalable}, which, however, still remain at the level of proposals. 
Moreover, recent theory works have shown that at low energies and even in far-from-equilibrium quench dynamics, QED can be approximately realized at small values of $S\lesssim4$ \cite{Zache2021achieving,Halimeh2021achieving}, which motivates pushing gauge-theory quantum-simulator technology to accommodate intermediate values of $S$ to probe the limit of lattice QED. In addition to their relevance to QED, $\mathrm{U}(1)$ QLMs at $S>1/2$ have been shown to also host exotic nonergodic regimes relevant to quantum many-body physics \cite{Desaules2022weak,Desaules2022prominent}. Achieving the capability to probe such phenomena on a quantum simulator can therefore provide deeper insights into ergodicity-breaking mechanisms in interacting quantum systems \cite{Rigol_review,Deutsch_review}.

Here, we present a mapping of $\mathrm{U}(1)$ QLMs onto bosonic superlattices at any value of $S$. We then propose an experimentally viable scheme for the realization of a large-scale spin-$1$ $\mathrm{U}(1)$ QLM on an extended Bose--Hubbard optical superlattice. Employing perturbation theory and MPS calculations, we demonstrate the fidelity of the mapping and the stability of gauge invariance over all accessible evolution times. We also showcase the versatility of our proposed quantum simulator by probing on it the (de)confinement of an electron--positron pair through tuning the gauge coupling.

The rest of the paper is organized as follows: In Sec.~\ref{sec:models}, we review spin-$S$ $\mathrm{U}(1)$ QLMs and outline their mapping onto bosonic superlattices. We provide an experimentally feasible concrete proposal for the implementation of a spin-$1$ $\mathrm{U}(1)$ QLM on an ultracold-atom quantum simulator in Sec.~\ref{sec:experiment}. Using matrix product state calculations, we benchmark in Sec.~\ref{sec:QuanchDynamics} the validity of the proposed quantum simulator through global quenches, and also probe on it the confinement--deconfinement transition. In Sec.~\ref{sec:GaugeProtection}, we provide analytic arguments to explain the robust stabilization of gauge violation on our proposed quantum simulator. We conclude and provide outlook in Sec.~\ref{sec:conc}. We supplement our work through various Appendices detailing our perturbation theory derivations (Appendix~\ref{app:PT}), providing details on our numerics (Appendix~\ref{app:MPS}), showcasing and benchmarking an alternative experimental architecture (Appendix~\ref{app:linear}), and comparing different QLM formulations (Appendix~\ref{app:idealQLM}).

\section{Spin-$S$ $\mathrm{U}(1)$ QLM and bosonic mapping}\label{sec:models}
Within QED, the gauge and electric fields are infinite-dimensional \cite{Zee_book}. For both experimental as well as theoretical purposes, however, it is convenient to adopt a quantum link formulation \cite{Chandrasekharan1997} of the QED Hamiltonian on a lattice \cite{Kogut1975,Carroll1976}, which results in the spin-$S$ $\mathrm{U}(1)$ QLM with Hamiltonian \cite{Chandrasekharan1997,Wiese_review,Hauke2013,Kasper2017}
\begin{align}\nonumber
    \hat{H}_\text{QLM}=&\frac{\kappa}{2\sqrt{S(S+1)}}\sum_{\ell=1}^{N-1}\big(\hat{\sigma}^-_\ell\hat{s}^+_{\ell,\ell+1}\hat{\sigma}^-_{\ell+1}+\text{H.c.}\big)\\\label{eq:QLM}
    &+\frac{\mu}{2}\sum_{\ell=1}^N\hat{\sigma}^z_{\ell}+\frac{g^2}{2}\sum_{\ell=1}^{N-1}\big(\hat{s}^z_{\ell,\ell+1}\big)^2,
\end{align}
where now the local gauge and electric fields on the link between sites $\ell$ and $\ell+1$ are represented by the spin-$S$ operators $\hat{s}^+_{\ell,\ell+1}/\sqrt{S(S+1)}$ and $\hat{s}^z_{\ell,\ell+1}$, respectively. The Pauli ladder operators $\hat{\sigma}^\pm_{\ell}$ act on the matter field on site $\ell$, where $\hat{\sigma}^z_\ell$ represents matter occupation, $\mu$ is the fermionic mass, $N$ is the total number of sites, $g$ is the gauge coupling, and $\kappa$ is the tunneling constant. We have set the lattice spacing to unity throughout. The generator of the $\mathrm{U}(1)$ gauge symmetry of Hamiltonian~\eqref{eq:QLM} is
\begin{align}\label{eq:G}
    \hat{G}_\ell=(-1)^\ell\bigg(\frac{\hat{\sigma}^z_\ell+\mathds{1}}{2}+\hat{s}^z_{\ell-1,\ell}+\hat{s}^z_{\ell,\ell+1}\bigg),
\end{align}
where $\big[\hat{H}_\text{QLM},\hat{G}_{\ell}\big]=\big[\hat{G}_\ell,\hat{G}_{\ell'}\big]=0,\,\forall\ell,\ell'$. The generator~\eqref{eq:G} can be viewed as a discretized version of Gauss's law.

\begin{figure}[t!!]
	\centering
	\includegraphics[width=\linewidth]{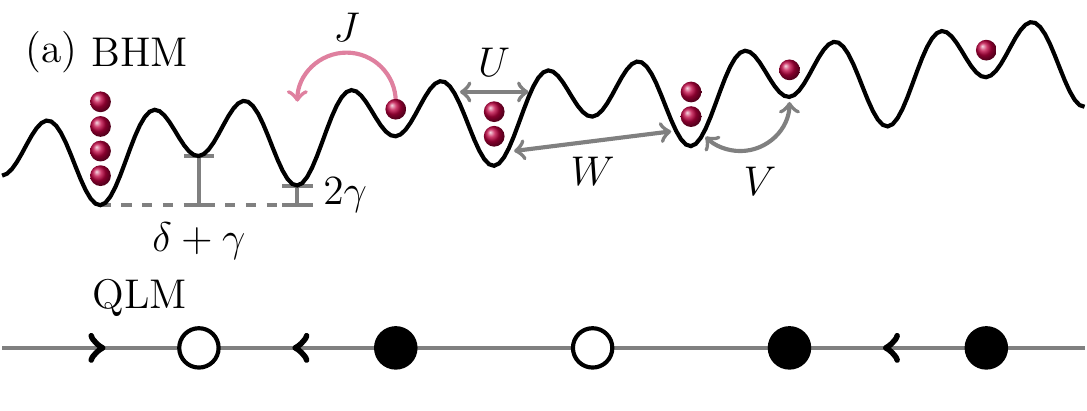}
	\includegraphics[width=\linewidth]{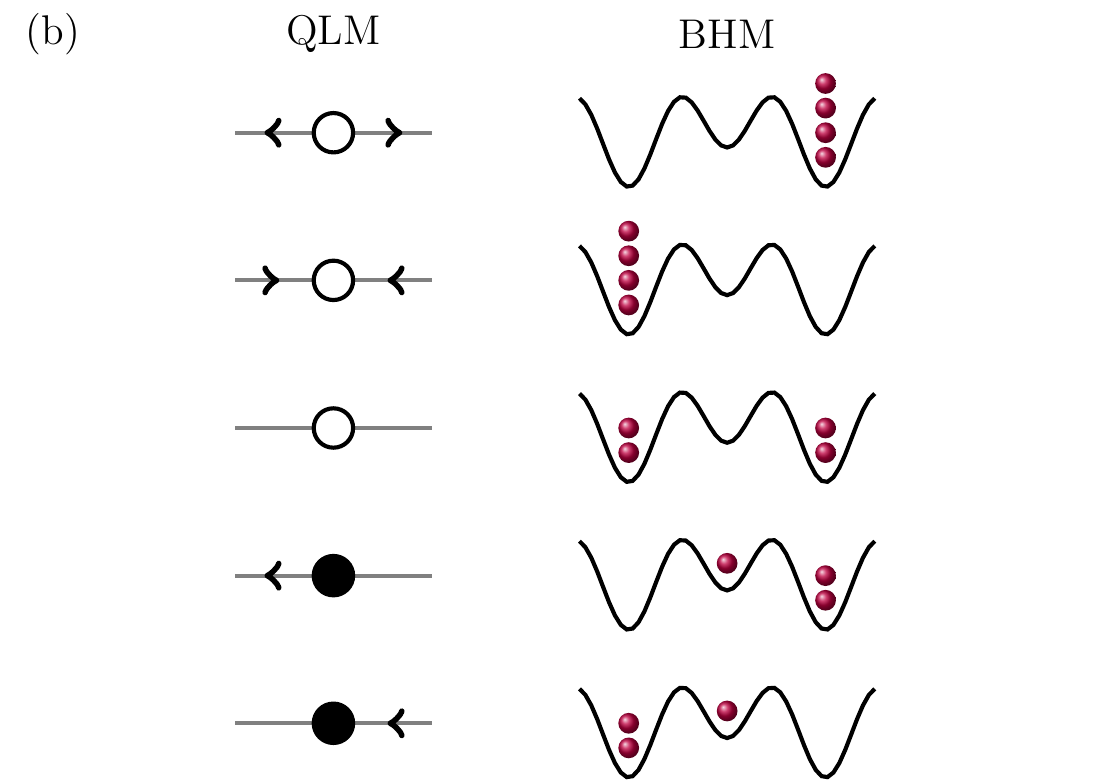}
	\includegraphics[width=\linewidth]{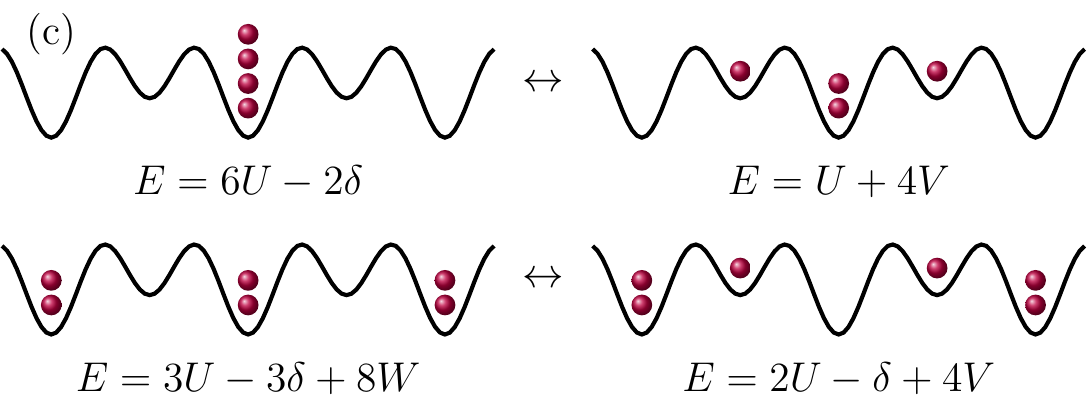}
	\caption{(Color online). (a) The extended Bose--Hubbard model on a tilted optical superlattice with spinless bosons, shown schematically for the case of the spin-$1$ $\mathrm{U}(1)$ QLM~\eqref{eq:tar} as the target theory. The tunneling strength is $J$, on-site interaction strength is $U$, staggering potential is $\delta$, the nearest-neighbor interaction is $V$, the next-nearest-neighbor interaction on the odd sites is $W$, and the tilted potential is $\gamma$. The shallow (deep) sites host matter (gauge) degrees of freedom. The tilted potential $\gamma$ enables the stabilization of the $\mathrm{U}(1)$ gauge invariance. (b) The configurations allowed by Gauss's law~\eqref{eq:G} shown in the QLM and BHM representations, where a leftward arrow on a link indicates an electric-field operator eigenvalue of $s^z=-1$ (represented by $0$ bosons), a rightward arrow represents $s^z=+1$ (represented by $4$ bosons), and an arrowless link denotes $s^z=0$ (represented by $2$ bosons). (c) The BHM parameters are determined by enforcing resonances between the shown allowed configurations.}
	\label{fig:mapping}
\end{figure}

We propose a mapping of the spin-$S$ quantum link model to a superlattice where even (odd) sites represent the matter sites (gauge links) of the gauge theory; see Fig.~\ref{fig:mapping}(a). On the even sites, we restrict the local Hilbert subspace to $\mathcal{H}_\mathrm{m}=\text{span}\{0,1\}$, while on odd sites to $\mathcal{H}_\mathrm{g}=\text{span}\{2n\},\,n\in\{0,1,\ldots,2S\}$. In terms of the electric flux eigenvalues $s^z$ on the odd sites, the allowed bosonic occupations $n_\text{g}=0,2,\ldots,4S-2,4S$ correspond to $s^z=-S,-S+1,\ldots,S-1,S$, respectively, i.e., $n_\text{g}=2(s^z+S)$; see Fig.~\ref{fig:mapping}(b) for the case of $S=1$. This corresponds to the mapping
\begin{subequations}\label{eq:mappings}
\begin{align}\label{eq:matter_mapping}
&\hat{\sigma}^+_\ell=\hat{\mathcal{P}}_\ell\hat{b}^\dagger_\ell\hat{\mathcal{P}}_\ell,\,\,\,\,\,\,\,\hat{\sigma}^z_\ell=\hat{\mathcal{P}}_\ell\big(2\hat{b}^\dagger_\ell\hat{b}_\ell-1\big)\hat{\mathcal{P}}_\ell,\\\label{eq:gauge_mapping}
&\hat{\tau}^+_{\ell,\ell+1}=\frac{1}{\sqrt{2S(2S+1)}}\hat{\mathcal{P}}_{\ell,\ell+1}\big(\hat{b}^\dagger_{\ell,\ell+1}\big)^2\hat{\mathcal{P}}_{\ell,\ell+1},\\\label{flux_mapping}
&\hat{s}^z_{\ell,\ell+1}=\frac{1}{4S}\hat{\mathcal{P}}_{\ell,\ell+1}\big(\hat{b}^\dagger_{\ell,\ell+1}\hat{b}_{\ell,\ell+1}-2S\big)\hat{\mathcal{P}}_{\ell,\ell+1},
\end{align}
\end{subequations}
where $\hat{b},\hat{b}^\dagger$ are local bosonic ladder operators, $\hat{\mathcal{P}}_\ell$ and $\hat{\mathcal{P}}_{\ell,\ell+1}$ are projectors onto the local Hilbert subspaces $\mathcal{H}_\mathrm{m}$ and $\mathcal{H}_\mathrm{g}$, respectively, and in Eq.~\eqref{eq:gauge_mapping} we have introduced a ``proto'' spin-$S$ representation of the gauge field in lieu of the rescaled actual spin-$S$ raising operator $\hat{s}^+_{\ell,\ell+1}/\sqrt{S(S+1)}$. Strictly speaking, a QLM is adequate to describe the low-energy subspace of lattice QED. From a particle-physics perspective, this means that the QLM representation is valid only for states $\ket{S,s^z}$ with $\lvert s^z\rvert\ll S$. As such, in the limit of $S\to\infty$, where the lattice QED limit is achieved, and for finite $s^z$, we have $\bra{S,\tilde{s}^z}\hat{s}^+\ket{S,s^z}/\sqrt{S(S+1)}=\delta_{\tilde{s}^z,s^z+1}\sqrt{S(S+1)-\tilde{s}^zs^z}/\sqrt{S(S+1)}\to1$. It is in this low-energy limit that one cannot resolve the difference between the infinite-dimensional gauge-field operator of lattice QED and $\hat{s}^+/\sqrt{S(S+1)}$. Similarly, in the regime of $S\gg1$ and a bosonic occupation $n_\text{g}\approx 2S-\alpha$ [where $\alpha\ll S$ and is even (odd) for (half-)integer $S$] on the gauge site, we have $\bra{n_\text{g}+2}\hat{\tau}^+\ket{n_\text{g}}\approx1$, and thus $\hat{\tau}^+$ will be indistinguishable from $\hat{s}^+/\sqrt{S(S+1)}$, and hence from the gauge-field operator of lattice QED, in the low-energy limit. 
These considerations are akin to the ones justifying the approximation of lattice QED through highly-occupied bosonic models \cite{Zohar2011,Yang:2017pra}. 
In what follows, we will use $\hat{\tau}^+$. A comparison to the dynamics under $\hat{s}^+$ is given in Appendix~\ref{app:idealQLM}, showing little qualitative difference between the two representations for relevant physical phenomena. 

\section{Ultracold-atom setup}\label{sec:experiment}
In the rest of this work, we focus on the case of $S=1$. In this section, we outline an experimentally feasible scheme for mapping the target QLM Hamiltonian
\begin{align}\nonumber
    \hat{H}_\text{tar}=&\frac{\kappa}{2}\sum_{\ell=1}^{N-1}\big(\hat{\sigma}^-_\ell\hat{\tau}^+_{\ell,\ell+1}\hat{\sigma}^-_{\ell+1}+\text{H.c.}\big)+\frac{\mu}{2}\sum_{\ell=1}^N\hat{\sigma}^z_{\ell}\\\label{eq:tar}
    &+\frac{g^2}{2}\sum_{\ell=1}^{N-1}\big(\hat{s}^z_{\ell,\ell+1}\big)^2,
\end{align}
onto the extended Bose--Hubbard model (BHM)
\begin{align}\nonumber
    \hat{H}_\text{BHM}=&-J\sum_{j=1}^{L-1}\big(\hat{b}_j^\dagger\hat{b}_{j+1}+\text{H.c.}\big)+\frac{U}{2}\sum_{j=1}^L\hat{n}_j\big(\hat{n}_j-1\big)\\\nonumber
    &+\sum_{j=1}^L\left[(-1)^j\frac{\delta}{2}+j\gamma\right]\hat{n}_j+V\sum_{j=1}^{L-1}\hat{n}_j\hat{n}_{j+1}\\\label{eq:eBHM}
    &+W\sum_{j=1}^{\frac{L}{2}-1}\hat{n}_{2j-1}\hat{n}_{2j+1}.
\end{align}
Here, $L=2N$ is the total number of bosonic sites $j$ with even (odd) $j$ representing a matter (gauge) site, $\hat{n}_j=\hat{b}_j^\dagger\hat{b}_j$ is the bosonic number operator, $U$ is the on-site interaction strength, $\delta$ is a staggered potential making even (odd) sites shallow (deep), and $\gamma$ is a lattice tilt.  
The microscopic model with target $S>1/2$ has more configurations available than $S=1/2$. To render the physical configurations resonant in the case of $S=1$, this requires a crucial further ingredient, namely interactions beyond on-site given by the nearest-neighbor interaction strength $V$ and  the strength of interactions between two consecutive odd sites $W$ (see Fig.~\ref{fig:mapping}).

These parameters are not independent, and they should be chosen such that the gauge-invariant processes between configurations allowed by Gauss's law are resonant; see Fig.~\ref{fig:mapping}(c). This leads to two equations with four unknowns:
\begin{subequations}\label{eq:underdetermined}
\begin{align}
    &5U-2\delta-4V\approx0,\\
    &U-2\delta-4V+8W\approx0.
\end{align}
\end{subequations}
This \textit{underdetermined} system of equations allows for a large range of possible values for the parameters, which can be scanned for experimental feasibility.

The parameters of Eqs.~\eqref{eq:QLM} and~\eqref{eq:eBHM} can be connected to each other by relating the processes in Fig.~\ref{fig:mapping}(c) in the QLM and BHM pictures and employing perturbation theory (up to second order), which leads to (see Appendix~\ref{app:PT} for derivational details)
\begin{subequations}
\begin{align}
    \mu&=-\frac{3}{2}U+\delta+2V-2W+\frac{16J^2\big(5U-6\delta\big)}{\big(5U-6\delta\big)^2-16\gamma^2},\\
    g^2&=4U-8W,\\    
    \kappa&=\frac{16\sqrt{6}J^2\big(2\delta-3U\big)}{(2\delta-3U\big)^2-16\gamma^2}.
\end{align}
\end{subequations}

Ultracold atoms trapped in optical lattices form a convenient quantum simulator for the realization of the target spin-$S$ $\mathrm{U}(1)$ QLM given by Eq.~\eqref{eq:tar}.
The envisioned system consists of ultracold dipolar bosonic atoms and a $1$D optical superlattice with a relatively small periodicity, described by an extended Bose--Hubbard model where a long-range-type interaction is included \cite{Dutta:2015,Baier:2016,Chomaz:2023}.

We suggest using a bichromatic superlattice to construct the $1$D trapping potential \cite{Folling:2007,Trotzky:2008}.
For the 1D linear case, the lattice potential can be written as $V_s\cos^2(4\pi x/\lambda) - V_l\cos^2(2\pi x/\lambda - \pi/4)$, where $V_{s,l}$ are the lattice depths of the `short' and the `long' lattice, respectively, and $\lambda$ is the wavelength of the `short' lattice. 
In such a configuration, the dynamics of the model can be realized by adjusting the lattice depth, thus tuning the tunnelling $J$ and the on-site interaction $U$, as well as the relative depth of the superlattice, thus tuning the energy offset $\delta$ \cite{Yang2020}. 
The lattice tilt $\gamma$ can be realized by a linear potential acting on the atoms, such as generated by the gravitational force or by a magnetic field acting on the atoms. In addition, the nearest- and next-nearest-neighbor interactions, $V$ and $W$, respectively can be realized by controlling the dipole-dipole interaction between highly magnetic atoms \cite{Chomaz:2023,Lahaye:2009,Griesmaier:2005,Lu:2011,Aikawa:2012}. 

To prepare the initial state with a special ordering, we propose driving the system from a superfluid state to a Mott-insulator regime in an optical superlattice \cite{Greiner2002,Spielman:2007,Bakr:2010,Sherson:2010, Yang:2020Science}. 
The overlapping of the intensity minima of the optical lattices creates a large occupation imbalance between neighboring lattice sites, which directs the atoms to occupy the `deep' lattice sites.
This allows us to engineer the atom occupation to satisfy Gauss's law of Eq.~\eqref{eq:G}, resulting in the system being tailored into the gauge-allowed subspace. For instance, we can generate the vacuum state denoted by $\ket{\ldots,2,0,2,0,2,0,\ldots}$.
The number fluctuations of this Mott-insulator state can be controlled by lowering the temperature of the system \cite{Yang:2020Science} or by implementing a state manipulation process \cite{Yang:2017pra,Greif:2013}. 

A crucial ingredient of the above setup are interactions beyond on-site. To judge the feasibility, we estimate the interaction strengths of bosonic Dysprosium atoms \cite{Lu:2011} in the type of optical superlattice proposed above.
When the short-lattice laser is set to a wavelength of $\lambda = 380\,\text{nm}$, the maximum nearest-neighbor dipole--dipole interaction in the repulsive side is around $V = 160\,\text{Hz}$, where we set Planck's constant to unity.
In a linear-type lattice, this interaction between atoms scales with $1/x^3$ over distance $x$, and therefore the next-nearest-neighbor interaction in a linear system has a strength of approximately $ W = V/2^3 = 20 \,\text{Hz}$.
Unfortunately, such a relation $V = 8W$ leads to unwanted terms appearing in second-order perturbation theory which restrict the dynamics of the gauge theory, as shown in Appendix~\ref{app:inhomog}.
These unwanted terms can be removed by setting $V = 2W$. However, obtaining such a ratio requires us to shorten the distance between relevant sites.
To achieve a controllable distance between the next-nearest-neighbor lattice sites, instead of a linear $1$D lattice, a zigzag ladder can be employed \cite{Greschner2013,zhang2015one,Cabedo2020,Roy2022Genuine,barbiero2022frustrated}. 
This approach enables us to enhance the interaction strength to approximately $W = 80\,\text{Hz}$, thus removing unwanted processes in second-order perturbation theory.

The dipolar interactions between atoms imposes constraints on the other energy scales such as $J$, $U$, $\delta$, and $\gamma$.
In the Hubbard regime, the corresponding parameter strengths can be reached by tuning the lattice confinement.
Therefore, one can realize the spin-$S$ model in a state-of-the-art extended Bose--Hubbard quantum simulator  while suppressing any gauge-violating dynamics.

\section{Quench dynamics}\label{sec:QuanchDynamics}
We now benchmark our proposed ultracold-atom quantum simulator by probing quench dynamics on it and comparing with the target QLM~\eqref{eq:tar}. For this, we employ MPS techniques \cite{Uli_review,Paeckel_review,mptoolkit} and perform the time evolution using the time-dependent variational principle (TDVP) \cite{Haegeman2011,Haegeman2016}. Details of the numerical implementation are provided in Appendix~\ref{app:MPS}. Our benchmarking will focus on the spatial averages of electric flux, chiral condensate, and gauge violation, 
\begin{subequations}
\begin{align}
    \mathcal{E}(t)&=\frac{1}{N-1}\sum_{\ell=1}^{N-1}(-1)^\ell\langle\hat{s}^z_{\ell,\ell+1}(t)\rangle, \label{eq:flux} \\
    \mathcal{C}(t)&=\frac{1}{2}+\frac{1}{2N}\sum_{\ell=1}^N\langle\hat{\sigma}^z_\ell(t)\rangle, \label{eq:cc} \\
    \eta(t)&=\sqrt{\frac{1}{N}\sum_{\ell=1}^N\langle\hat{G}_\ell^2(t)\rangle}, \label{eq:violation}
\end{align}
\end{subequations}
respectively. We seek very good quantitative agreement in the quench dynamics of the electric flux and chiral condensate between the BHM and QLM, which should correspond to a very small and controlled gauge violation over the entire time evolution.

Throughout this section, we set our parameters to \(J = 3\)\,Hz, \(\gamma = 7\)\,Hz, \(V = 160\)\,Hz, and \(W = 80\)\,Hz, which are experimentally feasible values in the zigzag scheme discussed in Sec.~\ref{sec:experiment}. Benchmarks of the linear architecture are included in Appendix~\ref{app:linear}.

\begin{figure}[t!!]
	\centering
	\includegraphics[width=\linewidth]{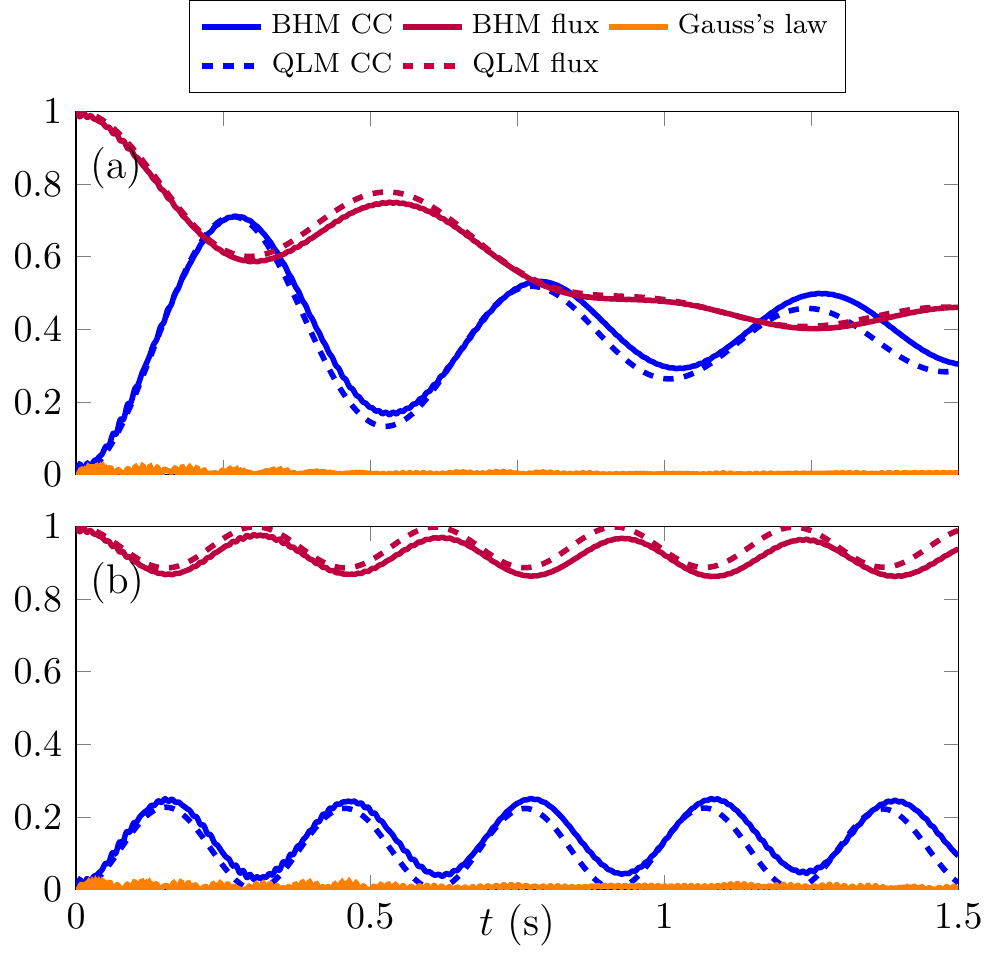}
	\caption{(Color online). Numerical simulations of a global quench of the extreme vacuum state in the BHM~\eqref{eq:eBHM} (solid lines) and target QLM~\eqref{eq:tar} (dashed). The values of the electric flux~\eqref{eq:flux} and chiral condensate~\eqref{eq:cc} show good agreement, and the gauge violation in the BHM~\eqref{eq:violation} remains on low levels throughout the entire time evolution.
    (a) Quench to \(U = 160\)\,Hz and \(\delta = 80\)\,Hz, which corresponds to \(\mu/\kappa \approx -0.41\) and \(g^2/\kappa = 0\) in the target model.
    (b) Quench to \(U = 159\)\,Hz and \(\delta = 78.5\)\,Hz, which corresponds to \(\mu/\kappa \approx -0.40\) and \(g^2/\kappa \approx 3.60\) in the target model.}
	\label{fig:quench-ev}
\end{figure}

\begin{figure}[t!!]
	\centering
	\includegraphics[width=\linewidth]{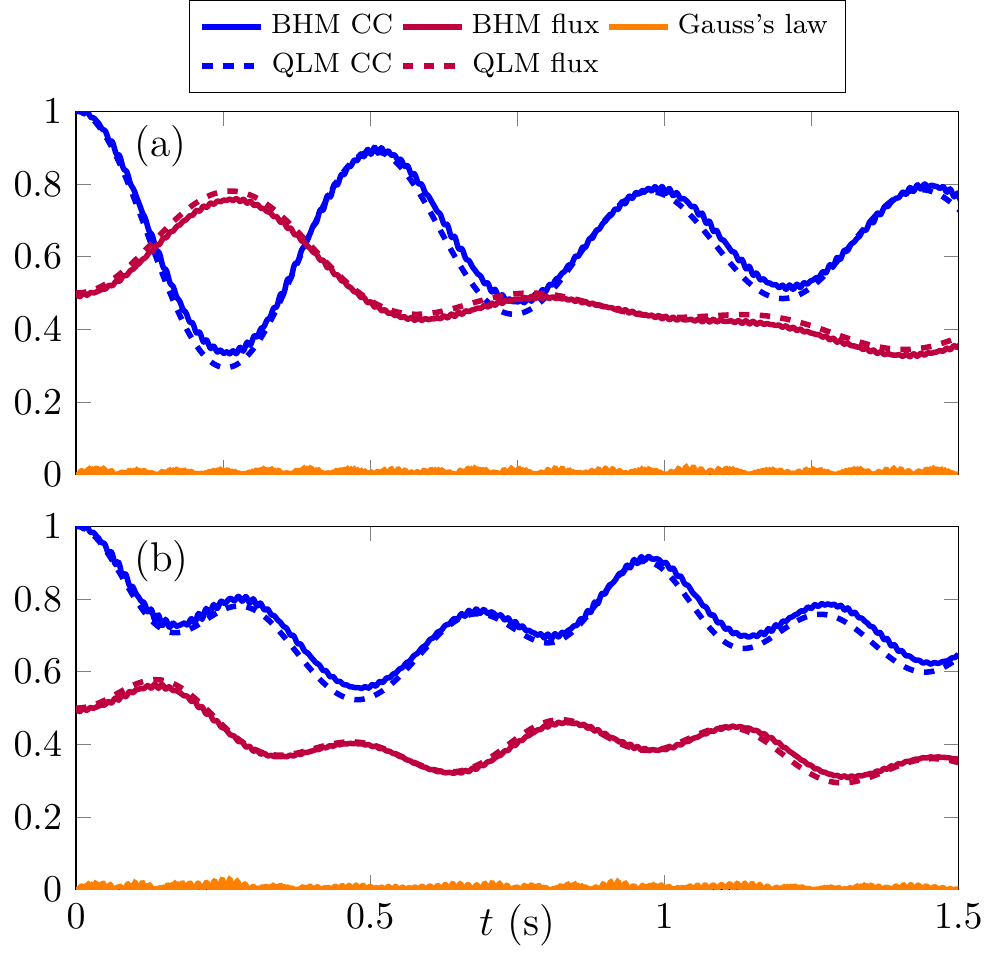}
	\caption{(Color online). Numerical simulations of a global quench of the charge-proliferated state in the BHM~\eqref{eq:eBHM} and target QLM~\eqref{eq:tar}. Again, the values of the electric flux~\eqref{eq:flux} and chiral condensate~\eqref{eq:cc} show good agreement, and the gauge violation in the BHM~\eqref{eq:violation} remains small.
    The quench parameters for each panel are the same as in the corresponding panels of Fig.~\ref{fig:quench-ev}.}
	\label{fig:quench-cp}
\end{figure}

We are interested in gauge-invariant translation-invariant states in the QLM that can be represented by an infinitely repeating unit cell of two matter sites and two gauge links $\lvert\sigma^z_\ell,s^z_{\ell,\ell+1},\sigma^z_{\ell+1},s^z_{\ell+1,\ell+2}\rangle$, where $\sigma^z_\ell$ and $s^z_{\ell,\ell+1}$ are the eigenvalues of $\hat{\sigma}^z_\ell$ and $\hat{s}^z_{\ell,\ell+1}$, respectively. In the bosonic mapping, this state is represented as $\ket{n_j,n_{j+1},n_{j+2},n_{j+3}}$, where $n_j$ is the eigenvalue of $\hat{n}_j$ of the optical superlattice, and $j$ is an even site representing a matter site in the QLM, and odd bosonic sites in the BHM represent gauge links in the QLM. We emphasize that our MPS calculations work directly in the thermodynamic limit for these states, and they are infinite in spatial extent (see Appendix~\ref{app:MPS}). 

In particular, we consider the following three types of initial states (``vacua''). The ``highly excited'' or ``extreme vacua'' of the QLM, given by $\ket{-1,\pm1,-1,\mp1}$ (bosonic mapping: $\ket{0,4,0,0}$ and $\ket{0,0,0,4}$, respectively), corresponding to absence of charges and maximal allowed values of electric field (for $S=1$); the ``middle vacuum'' $\ket{-1,0,-1,0}$ (bosonic mapping: $\ket{0,2,0,2}$), corresponding to absence of charged particles and vanishing electric field; and the charge-proliferated states $\ket{+1,0,+1,-1}$ and $\ket{+1,-1,+1,0}$ (bosonic mapping: $\ket{1,2,1,0}$ and $\ket{1,0,1,2}$), corresponding to presence of charges with electric fields between neighboring pairs of particles; see Fig.~\ref{fig:mapping}(b) for the mapping between the QLM and BHM pictures.

\subsection{Extreme vacuum}\label{sec:EV}
To begin with, we consider the extreme vacuum state $\ket{-1,1,-1,-1}$, which in the bosonic picture is $\ket{0,4,0,0}$. We first consider a quench at $U=160$\,Hz and $\delta=80$\,Hz, which corresponds to $\mu/\kappa\approx-0.41$ and $g^2/\kappa=0$ in the target QLM. The corresponding quench dynamics of the electric flux, chiral condensate, and gauge violation are shown in Fig.~\ref{fig:quench-ev}(a), with solid (dashed) lines corresponding to the BHM (QLM). We see very good quantitative agreement in the quench dynamics of the electric flux and chiral condensate between the BHM and QLM, which is a testament to the faithfulness of our proposed quantum simulator. Indeed, the gauge violation over the whole duration of the dynamics, roughly $1.5$ seconds, is very well suppressed and never exceeds $3\%$.

As a second benchmark case, we quench the extreme vacuum to the model parameters $U=159$\,Hz and $\delta=78.5$\,Hz, which corresponds to $\mu/\kappa\approx-0.40$ and $g^2/\kappa\approx3.60$. The corresponding quench dynamics is shown in Fig.~\ref{fig:quench-ev}(b). Again, we find very good quantitative agreement in the dynamics of the electric flux and chiral condensate between the BHM and QLM, and the gauge violation is very well suppressed over all accessible evolution times.

From a phenomenological point of view, we see that the dynamics at the larger value of $g^2/\kappa$ is significantly constrained. We will investigate this behavior in more detail below by probing the fate of an electron--positron pair depending on the strength of the gauge coupling.

\begin{figure}[t!!]
	\centering
	\includegraphics[width=\linewidth]{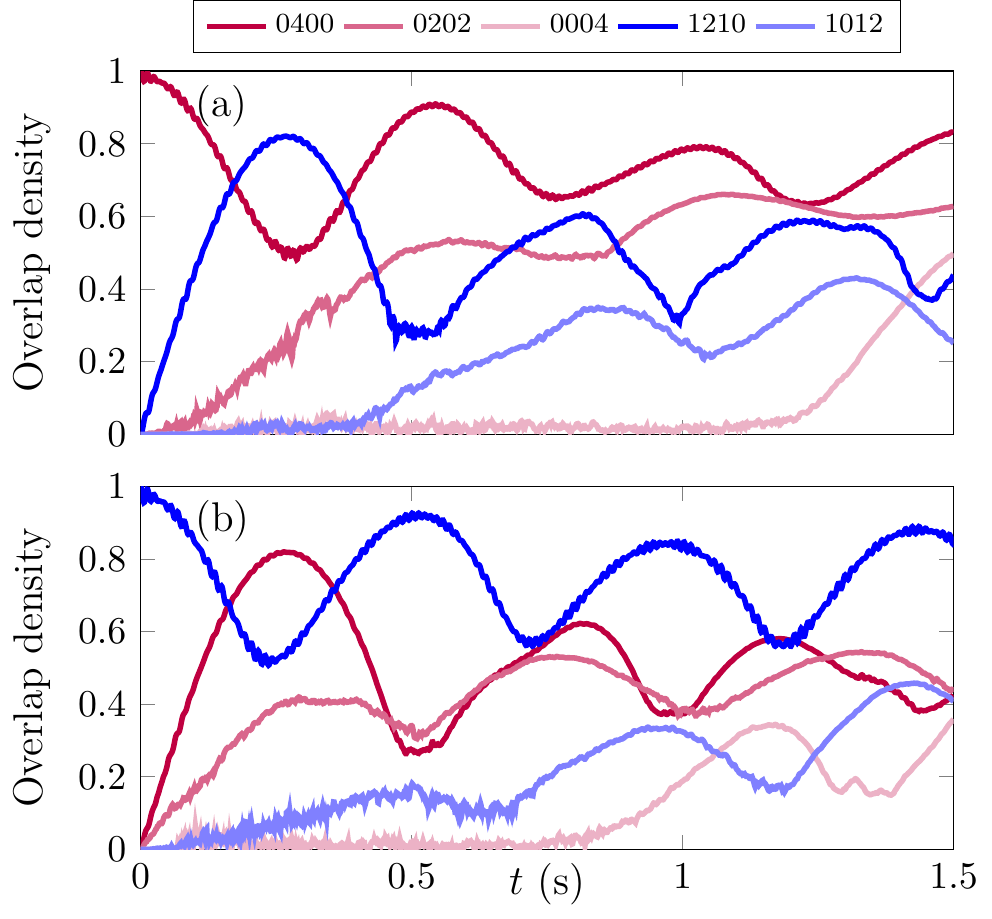}
	\caption{(Color online). Plot of the overlap density (per four site unit cell) of the time evolved states of the global quenches in the BHM with the three vacuum states (shown in shades of red) and two charge-proliferated states (in shades of blue).
    The time evolution results shown correspond to Fig.~\ref{fig:quench-ev}(a) in panel (a) and Fig.~\ref{fig:quench-cp}(a) in panel (b).}
	\label{fig:overlaps}
\end{figure}

\begin{figure*}[t!!]
	\centering
	\includegraphics[width=\linewidth]{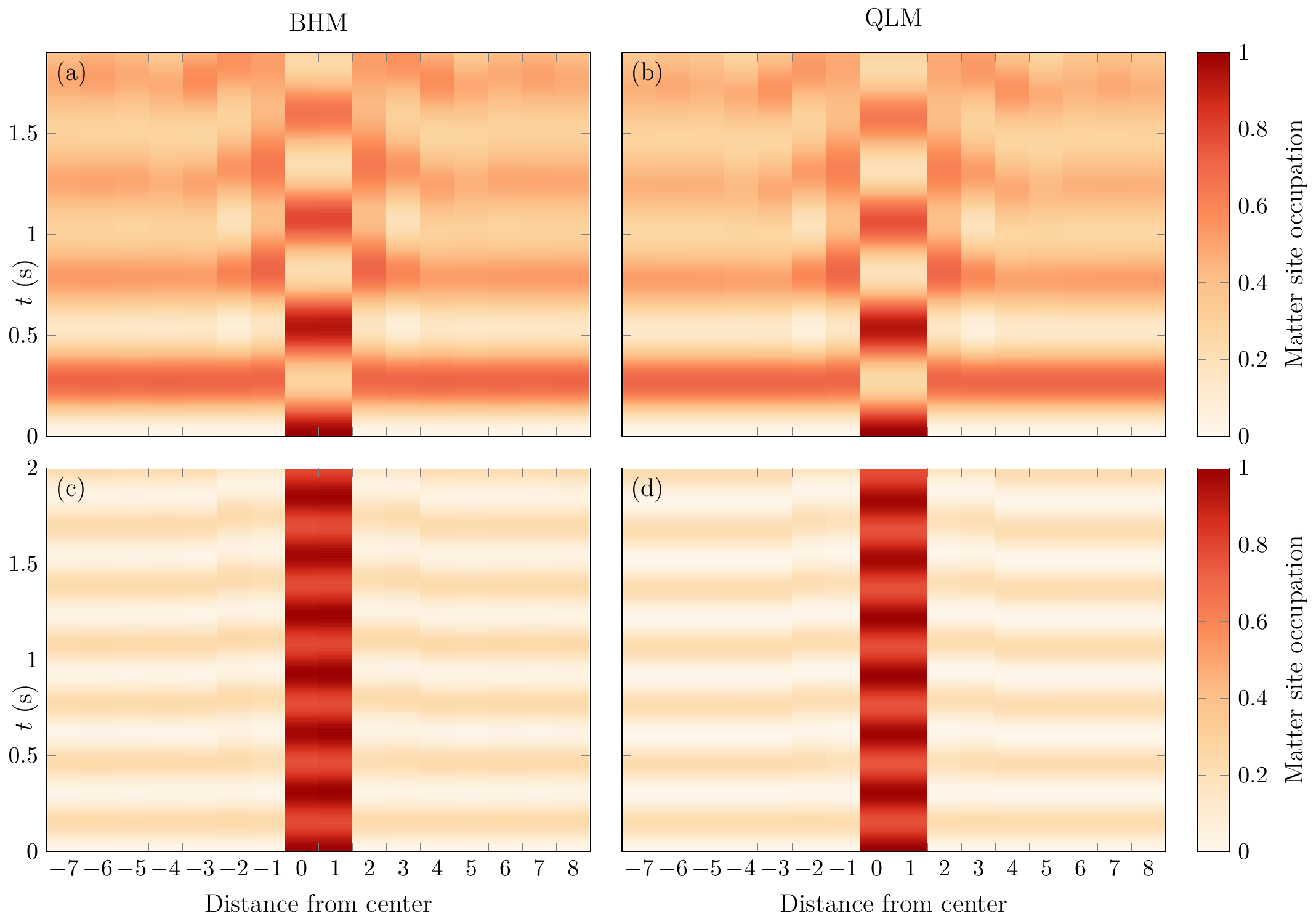}
	\caption{(Color online). Numerical simulations of an electron--positron pair on top of the extreme vacuum in the BHM~\eqref{eq:eBHM} (panels (a) and (c)) and the target QLM~\eqref{eq:tar} (panels (b) and (d)).
    (a,b) Quench to \(U = 160\)\,Hz and \(\delta = 80\)\,Hz, which correspond to \(\mu/\kappa \approx -0.41\) and \(g^2/\kappa = 0\) in the target model.
	(c,d) Quench to \(U = 159\)\,Hz and \(\delta = 78.5\)\,Hz, which correspond to \(\mu/\kappa \approx -0.40\) and \(g^2/\kappa \approx 3.60\) in the target model.}
	\label{fig:pair}
\end{figure*}

\subsection{Charge-proliferated state}\label{sec:CP}
Next, we consider as initial state the charge-proliferated state $\ket{+1,0,+1,-1}$, which in the bosonic picture is $\ket{1,2,1,0}$. We repeat the same quenches employed in Sec.~\ref{sec:EV} for the extreme vacuum, and the corresponding quench dynamics are presented in Fig.~\ref{fig:quench-cp}. In both cases, we again see very robust quantitative agreement in the quench dynamics of the electric flux and chiral condensate between the BHM and QLM up to all investigated evolution times. The proposed BHM quantum simulator further demonstrates its faithfulness by exhibiting a very suppressed gauge violation throughout the entire dynamics.

It is interesting to calculate the overlap of the time-evolved wave function with the three vacua and two charge-proliferated states of the target QLM for quenches considered in Secs.~\ref{sec:EV} and~\ref{sec:CP}. We show this in Fig.~\ref{fig:overlaps}(a,b) for the quenches of Figs.~\ref{fig:quench-ev}(a) and~\ref{fig:quench-cp}(a), respectively. The quench dynamics shows significant overlap of the wave function with all these five product states, signifying the system indeed takes full advantage of the additional configurations available thanks to the effective spin-$1$ representation of the gauge and electric fields.

\subsection{Electron--positron pair and confinement}
Confinement is one of the most intriguing phenomena of gauge theories, and there have been several proposals to probe confinement in quantum simulators of gauge theories for spin-$1/2$ $\mathrm{U}(1)$ QLMs involving the implementation of a topological $\theta$-angle \cite{Surace2020,Halimeh2022tuning,Cheng2022tunable}. Whereas in the latter the gauge-coupling term $(g^2/2)\sum_\ell(\hat{s}^z_{\ell,\ell+1})^2$ is an inconsequential energy constant since $(\hat{s}^z_{\ell,\ell+1})^2=\mathds{1}/4$ for $S=1/2$, this term is dynamic in the case of $S=1$, and is known to cause confinement at strong gauge coupling (i.e., when $g$ is large) \cite{Chandrasekharan1999}. QLMs with integer values of $S$ resemble the Wilson-Kogut-Susskind formulation of $\mathrm{U}(1)$ lattice gauge theories \cite{Kogut1975}, in which confinement in the strong-coupling regime has been studied \cite{Byrnes2002}. The gauge-coupling term becomes equivalent to an energetic constraint that penalizes large deviations from the original configuration of flux strings. Indeed, at large values of $\lvert g^2/\kappa\rvert\gg1$, the gauge-coupling term can be thought of as inducing coherent quantum Zeno dynamics \cite{facchi2002quantum,facchi2004unification,facchi2009quantum,burgarth2019generalized} that constrains the time evolution in a small subspace of the total Gauss's law sector.

In order to properly probe (de)confinement in our target QLM, we turn to a paradigmatic test state and prepare an electron--positron pair on top of the extreme vacuum considered in Sec.~\ref{sec:EV}. This state breaks translation invariance, and as such we cannot employ infinite MPS, but rather we restrict ourselves to a finite system of $N=40$ matter sites (see Appendix~\ref{app:MPS}).

Quenching this pair state with the BHM model at $U=2\delta=160$\,Hz, which corresponds to $\mu/\kappa\approx-0.41$ and $g^2/\kappa=0$ in the target QLM, we find a rapid spread of the pair indicating deconfinement, as shown in the matter density profile in Fig.~\ref{fig:pair}(a,b) for the BHM and target QLM, respectively. Note how the quantitative agreement in the dynamics between both models is excellent over all investigated evolution times. The gauge violation in the BHM simulations is well suppressed and never exceeds $3\%$, as with the quenches of the translation-invariant states.

We now probe the dynamics of the electron--positron pair at a large value of the gauge coupling $g$. We quench the pair state with the BHM model at $U=159$\,Hz and $\delta=78.5$\,Hz, which corresponds to $\mu/\kappa\approx-0.40$ and $g^2/\kappa\approx3.60$ in the target QLM. The resulting quench dynamics, shown in Fig.~\ref{fig:pair}(c,d) for the BHM and QLM, respectively, shows fundamentally different behavior from the case of $g^2=0$. Indeed, the electron--positron pair appears to be bound and immobile for all accessible evolution times in MPS, which is a clear signature of confinement. Also in this case, the quantitative agreement in the quench dynamics between the BHM and QLM is excellent, with the gauge violation in the BHM simulation being well controlled.

Furthermore, this is the (de)confinement behavior expected in the ``ideal'' spin-$1$ $\mathrm{U}(1)$ QLM~\eqref{eq:QLM} \cite{Chandrasekharan1999}, and, indeed, the corresponding dynamics for the same quench in the latter is qualitatively similar to that of Fig.~\ref{fig:pair}, as shown in Fig.~\ref{fig:pair-ideal} in Appendix~\ref{app:idealQLM}.

\section{Linear gauge protection}\label{sec:GaugeProtection}
It would be instructive to get a better understanding of the excellent suppression of gauge violations that we find over all investigated evolution times in our quench dynamics. As already mentioned, the tunneling term in Hamiltonian~\eqref{eq:eBHM} is the perturbation that breaks gauge invariance, although it induces effective gauge-theory dynamics due to a large energetic penalty on processes driving the dynamics away from the physical sector. To illustrate this, let us consider the part of Hamiltonian~\eqref{eq:eBHM} that is diagonal in the particle-number basis, and write it using QLM spatial indexing,
\begin{align}\nonumber
    \hat{H}_\text{diag}=\sum_{\ell}\bigg\{&\frac{U}{2}\Big[\hat{n}_\ell\big(\hat{n}_\ell-1\big)+\hat{n}_{\ell,\ell+1}\big(\hat{n}_{\ell,\ell+1}-1\big)\Big]\\\nonumber
    &-\delta\hat{n}_{\ell,\ell+1}+2(-1)^\ell\ell\gamma\hat{G}_\ell+W\hat{n}_{\ell-1,\ell}\hat{n}_{\ell,\ell+1}\\\label{eq:diag}
    &+V\big[\hat{n}_{\ell-1,\ell}\hat{n}_\ell+\hat{n}_\ell\hat{n}_{\ell,\ell+1}\big]\bigg\},
\end{align}
up to an inconsequential energetic constant, where we have rewritten the generator~\eqref{eq:G} of Gauss's law in the bosonic basis as
\begin{align}
    \hat{G}_\ell=(-1)^\ell\bigg(\hat{n}_\ell+\frac{\hat{n}_{\ell-1,\ell}+\hat{n}_{\ell,\ell+1}}{2}-2S\bigg).
\end{align}
We can identify in Eq.~\eqref{eq:diag} a linear gauge protection term $\hat{H}_G=\sum_\ell2(-1)^\ell\ell\gamma\hat{G}_\ell$, which has been shown to stabilize gauge invariance up to all numerically accessible evolution times \cite{Halimeh2020e,Lang2022SGP}. This explains the excellent stabilization of gauge invariance in all of our results. The other diagonal terms in the BHM are needed to ensure that the accessible states in perturbation theory correspond to physical states in the QLM, as detailed in Appendix~\ref{app:PT}.

\section{Conclusion and outlook}\label{sec:conc}
We have presented a general mapping of spin-$S$ $\mathrm{U}(1)$ quantum link models onto bosonic ultracold atoms on an optical superlattice. In keeping with experimental feasibility on current ultracold-atom quantum technologies, we have proposed an extended single-species Bose--Hubbard quantum simulator of a spin-$1$ $\mathrm{U}(1)$ quantum link model. Using perturbation theory, we have derived the exact relations between the parameters of the quantum simulator and the quantum link model.

Using matrix product state techniques and the time-dependent variation principle, we benchmarked the quench dynamics of local observables in the quantum link model with that of the quantum simulator, showing great quantitative agreement up to all accessible evolution times. In all cases, the gauge violation was strongly suppressed and controlled over all investigated evolution times. We showed how this was a result of a linear gauge protection term naturally arising in our mapping.

We demonstrated the ability of our quantum simulator to probe relevant high-energy phenomena by calculating the quench dynamics of an electron--positron pair, showing a confinement--deconfinement transition by tuning the gauge coupling $g$.

Our work opens the door towards larger-spin representations of quantum link model regularizations of quantum electrodynamics on modern quantum-simulator platforms, and is amenable to several extensions. For example, a topological $\theta$-term can be implemented by adding an additional staggering to the gauge sites in the optical superlattice \cite{Halimeh2022tuning,Cheng2022tunable}. Furthermore, it is possible to extend our setup to $2+1$D along the lines of recent proposals \cite{osborne2022largescale,surace2023abinitio}.

\begin{acknowledgments}
The authors are grateful to the groups of Jian-Wei Pan and Zhen-Sheng Yuan for work on related projects. J.C.H.~is grateful to Debasish Banerjee and Guo-Xian Su for stimulating discussions. B.Y.~acknowledges support from National Key R\&D Program of China (Grant No 2022YFA1405800), NNSFC (Grant No 12274199) and the Stable Support Plan Program of Shenzhen Natural Science Fund (Grant No 20220815092422002). J.C.H.~acknowledges funding within the QuantERA II Programme that has received funding from the European Union’s Horizon 2020 research and innovation programme under Grand Agreement No 101017733, support by the QuantERA grant DYNAMITE, by the Deutsche Forschungsgemeinschaft (DFG, German Research Foundation) under project number 499183856, funding by the Deutsche Forschungsgemeinschaft (DFG, German Research Foundation) under Germany's Excellence Strategy -- EXC-2111 -- 390814868, and funding from the European Research Council (ERC) under the European Union’s Horizon 2020 research and innovation programm (Grant Agreement no 948141) — ERC Starting Grant SimUcQuam.
P.H.~acknowledges funding from Provincia Autonoma di Trento, and by Q@TN, the joint lab between University of Trento, FBK-Fondazione Bruno Kessler, INFN-National Institute for Nuclear Physics, and CNR-National Research Council. This project has received funding from the European Research Council (ERC) under the European Union’s Horizon $2020$ research and innovation programme (grant agreement No $804305$). Funded by the European Union under Horizon Europe Programme - Grant Agreement 101080086 — NeQST. Views and opinions expressed are however those of the author(s) only and do not necessarily reflect those of the European Union or European Climate, Infrastructure and Environment Executive Agency (CINEA). Neither the European Union nor the granting authority can be held responsible for them.
I.P.M.~acknowledges support from the Australian Research Council (ARC) Discovery Project Grants No.~DP190101515 and DP200103760.
Numerical simulations were performed on The University of Queensland's School of Mathematics and Physics Core Computing Facility \texttt{getafix}.

\end{acknowledgments}

\appendix
\section{Perturbation theory}\label{app:PT}
To determine the values of the tunneling strength \(\kappa\) and the diagonal terms (\(\mu\) and \(g\)) in the target QLM for our BHM setup, we only need to consider translation-invariant states with a four-site unit cell in the BHM (which corresponds to two matter sites and two gauge links in the QLM mapping).
This is sufficient to describe all possible homogeneous gauge-invariant states and the processes that couple them.
Since our BHM setup has interaction terms that affect inhomogeneous states nontrivially, we will examine the effect this has on the target QLM Hamiltonian afterwards.

\subsection{Homogeneous states}
Let us consider the possible states for a translation-invariant unit cell of four sites in the BHM, which corresponds to two matter sites in the QLM mapping.
We have a total of four bosons per four-site unit cell.
We only consider the states corresponding to gauge-invariant states in the QLM mapping and states which can be obtained from these states by hopping a single boson%
\footnote{In the regime where \(\gamma \gg J^2/U\), second-order processes where bosons tunnel to next-nearest-neighboring sites will be suppressed, so we do not consider them here.}
(for brevity, we will only consider one of the two extreme vacua and charge-proliferated states, since the other ones do not provide any additional information).
We label these states as \(\ket{i}\) for \(i = 1,\ldots,11\): their boson configurations (where the first site in the unit cell corresponds to a matter site) and energy densities are%
\footnote{Since the density of the tilt term in the Hamiltonian \(\sum_{j=1}^L j \gamma \hat{n}_j\) is divergent in the thermodynamic limit, we cannot assign an energy density for this term by inspecting the state just by itself. 
Since we are only considering states that are connected by hopping a single boson left or right, which will change the energy density by \(-\gamma\) and \(\gamma\) respectively, we set the tilt energy density for the state \(\ket{1}\) to be zero, and then the tilt energy density for the other states will depend on how many bosons need to hop left or right to arrive there from \(\ket{1}\).}
\begin{subequations}
\begin{align}
    \ket{1} &= \ket{0400}, & E_1 &= 6U - 2\delta, \\
    \ket{2} &= \ket{1300}, & E_2 &= 3U - \delta + 3V - \gamma, \\
    \ket{3} &= \ket{0310}, & E_3 &= 3U - \delta + 3V + \gamma, \\
    \ket{4} &= \ket{1210}, & E_4 &= U + 4V, \\
    \ket{5} &= \ket{2110}, & E_5 &= U + \delta + 3V - \gamma, \\
    \ket{6} &= \ket{1120}, & E_6 &= U + \delta + 3V + \gamma, \\
    \ket{7} &= \ket{0211}, & E_7 &= U - \delta + 3V + 4W - \gamma, \\
    \ket{8} &= \ket{1201}, & E_8 &= U - \delta + 3V + 4W + \gamma, \\
    \ket{9} &= \ket{0202}, & E_9 &= 2U - 2\delta + 8W, \\
    \ket{10} &= \ket{1102}, & E_{10} &= E_7, \\
    \ket{11} &= \ket{0112}, & E_{11} &= E_8.
\end{align}
\end{subequations}
The nonzero off-diagonal elements of the Hamiltonian \(\hat{H} = \hat{H}_\text{BHM}\) are
\begin{subequations}
\begin{align}
    \mel{1}{\hat{H}}{2} = \mel{1}{\hat{H}}{3} &= -2J, \\
    \mel{2}{\hat{H}}{4} = \mel{3}{\hat{H}}{4} &= -\sqrt{3}J, \\
    \mel{4}{\hat{H}}{5} = \mel{4}{\hat{H}}{6} &= -2J, \\
    \mel{4}{\hat{H}}{7} = \mel{4}{\hat{H}}{8} &= -J, \\
    \mel{i}{\hat{H}}{9} &= -\sqrt{2}J, \qquad i = 7,8,10,11,
\end{align}
\end{subequations}
and their conjugate elements, which are the same (e.g., \(\mel{2}{\hat{H}}{1} = \mel{1}{\hat{H}}{2}\)).

We now employ a Schrieffer--Wolff transformation on the Hamiltonian, which we rewrite in the form
\begin{equation}
    \hat{H} = \hat{H}_0 + \hat{H}_1,
\end{equation}
where \(\hat{H}_0\) contains all of the diagonal elements and \(\hat{H}_1\) contains all of the off-diagonal elements.
We then wish to find the transformed effective Hamiltonian
\begin{align}
    \hat{H}_\text{eff} = \mathrm{e}^{\mathrm{i} \hat{S}} \hat{H} \mathrm{e}^{-\mathrm{i} \hat{S}},
\end{align}
where \(\hat{S}\) is a small-valued Hermitian matrix that we need to find.
By the Baker--Campbell--Hausdorff formula, we have that
\begin{align}
    \hat{H}_\text{eff} &= \hat{H} + \mathrm{i} [\hat{S},\hat{H}] - \frac{1}{2} [\hat{S},[\hat{S},\hat{H}]] + \mathcal{O}(\hat{S}^3).
\end{align}
We neglect the terms \(\mathcal{O}(\hat{S}^3)\), and define \(\hat{S}\) such that
\begin{align}\label{eq:S}
    [\hat{S},\hat{H}_0] = \mathrm{i} \hat{H}_1,
\end{align}
which allows us to obtain
\begin{align}\label{eq:H-eff}
    \hat{H}_\text{eff} \approx \hat{H}_0 + \frac{1}{2} \mathrm{i} [\hat{S}, \hat{H}_1].
\end{align}

We can find the matrix elements of \(\hat{S}\) in the basis \(\{\ket{i}\}_{i=1,\ldots,9}\) by using our definition of \(\hat{S}\)~\eqref{eq:S}
and using the fact that \(\hat{H}_0\) is diagonal
\begin{equation}\label{eq:S-elements}
    \mel{i}{\hat{S}}{j} = \mathrm{i} \frac{\mel{i}{\hat{H}_1}{j}}{E_j - E_i},
\end{equation}
where \(E_i \equiv \mel{i}{\hat{H}_0}{i}\).
The matrix elements of the effective Hamiltonian \eqref{eq:H-eff} in the basis \(\{\ket{i}\}_{i=1,\ldots,9}\) are thus
\begin{multline}\label{eq:H-eff-elements-final}
    \mel{i}{\hat{H}_\text{eff}}{j} = \delta_{ij} E_i \\
    + \frac{1}{2} \sum_k \left( \frac{1}{E_i - E_k} + \frac{1}{E_j - E_k} \right) \mel{i}{\hat{H}_1}{k} \mel{k}{\hat{H}_1}{j}.
\end{multline}

Note that our assumption that \(\hat{S}\) was small-valued in order to neglect terms \(\mathcal{O}(\hat{S}^3)\) implies that we need \(\lvert\mel{i}{\hat{H}_1}{j}\rvert \ll \lvert E_j - E_i\rvert\) by Eq.~\eqref{eq:S-elements}.
That is, we require the energies of the gauge-invariant states \(\{\ket{1},\ket{4},\ket{9}\}\) to be well-separated from the energies of the other states to which they are coupled by \(\hat{H}_1\).
This condition is generally satisfied, although care needs to be taken particularly with the choice of \(\gamma\) to make sure such resonances do not occur.
In this regime, we can truncate the basis of the effective Hamiltonian to the gauge-invariant states, and calculate their matrix elements as follows (where the labels above each term show which state is responsible for it, and the notation \((\gamma \rightarrow -\gamma)\) is an abbreviation for the previous terms but with \(\gamma\) replaced with \(-\gamma\)):
\newpage
\begin{widetext}
\begin{subequations}
\begin{align}
    \mel{1}{\hat{H}_\text{eff}}{1} &= 6U - 2\delta + 4J^2 \bigg[ \overbrace{\frac{1}{3U - \delta - 3V + \gamma}}^{\text{via}~\ket{2}} + \overbrace{(\gamma \rightarrow -\gamma)}^{\text{via}~\ket{3}} \bigg], \\
    \mel{1}{\hat{H}_\text{eff}}{4} &= \sqrt{3} J^2 \bigg[ \overbrace{\frac{1}{3U - \delta - 3V + \gamma} + \frac{1}{-2U + \delta + V + \gamma}}^{\text{via}~\ket{2}} + \overbrace{(\gamma \rightarrow -\gamma)}^{\text{via}~\ket{3}} \bigg], \\
    \mel{4}{\hat{H}_\text{eff}}{4} &= U + 4V + J^2 \bigg[ \overbrace{\frac{3}{-2U + \delta + V + \gamma}}^{\text{via}~\ket{2}} + \overbrace{\frac{1}{\delta + V - 4W + \gamma}}^{\text{via}~\ket{7}} + \overbrace{\frac{4}{-\delta + V + \gamma}}^{\text{via}~\ket{5}} + \overbrace{(\gamma \rightarrow -\gamma)}^{\text{via}~\ket{3},\ket{6},\ket{8}} \bigg], \\
    \mel{4}{\hat{H}_\text{eff}}{9} &= \frac{\sqrt{2}J^2}{2} \bigg[ \overbrace{\frac{1}{\delta + V - 4W + \gamma} + \frac{1}{U - \delta - 3V + 4W + \gamma}}^{\text{via}~\ket{7}} + \overbrace{(\gamma \rightarrow -\gamma)}^{\text{via}~\ket{8}} \bigg], \\
    \mel{9}{\hat{H}_\text{eff}}{9} &= 2U - 2\delta + 8W + 4J^2 \bigg[ \overbrace{\frac{1}{U - \delta - 3V + 4W + \gamma}}^{\text{via}~\ket{7},\ket{10}} + \overbrace{(\gamma \rightarrow -\gamma)}^{\text{via}~\ket{8},\ket{11}} \bigg].
\end{align}
\end{subequations}
\end{widetext}
Close to resonance~\eqref{eq:underdetermined}, we have that
\begin{subequations}
\begin{align}
    V &\approx \frac{5U}{4} - \frac{\delta}{2}, \label{eq:resonance1}\\
    4W &\approx -\frac{U}{2} + \delta + 2V \approx 2U. \label{eq:resonance2}
\end{align}
\end{subequations}
By using these conditions, we can obtain simplified expressions for the parameters in the target QLM as
\begin{subequations}
\begin{align}
    g^2 &= \mel{1}{\hat{H}_\text{eff}}{1} - \mel{9}{\hat{H}_\text{eff}}{9} = 4U - 8W, \label{eq:g2}\\
    \mu &= \frac{1}{2} \left[ \mel{4}{\hat{H}_\text{eff}}{4} - \mel{9}{\hat{H}_\text{eff}}{9} - \frac{g^2}{2} \right] \nonumber\\
    &= -\frac{3}{2} U + \delta + 2V - 2W \nonumber\\
    &\quad + 2J^2 \left[ \frac{1}{5U/4 - 3\delta/2 + \gamma} + (\gamma \rightarrow -\gamma) \right], \label{eq:mu}\\
    \kappa &= \frac{2\sqrt{6}}{2\sqrt{3}}\mel{1}{\hat{H}_\text{eff}}{4} = \frac{2\sqrt{6}}{\sqrt{2}} \mel{4}{\hat{H}_\text{eff}}{9} \nonumber\\
    &= 2\sqrt{6} J^2 \left[ \frac{1}{-3U/4 + \delta/2 + \gamma} + (\gamma \rightarrow -\gamma) \right].
\end{align}
\end{subequations}

\subsection{Inhomogeneous states}\label{app:inhomog}
In the previous section, we only dealt with homogeneous gauge-invariant states where the matter sites were either all occupied or all unoccupied.
Since our BHM setup has first- and second-neighbor interaction terms, the renormalized energies of inhomogeneous configurations will differ from the sum of energies obtained from each individual site using the terms in the previous section.

\begin{figure}[t!!]
	\centering
	\includegraphics[width=\linewidth]{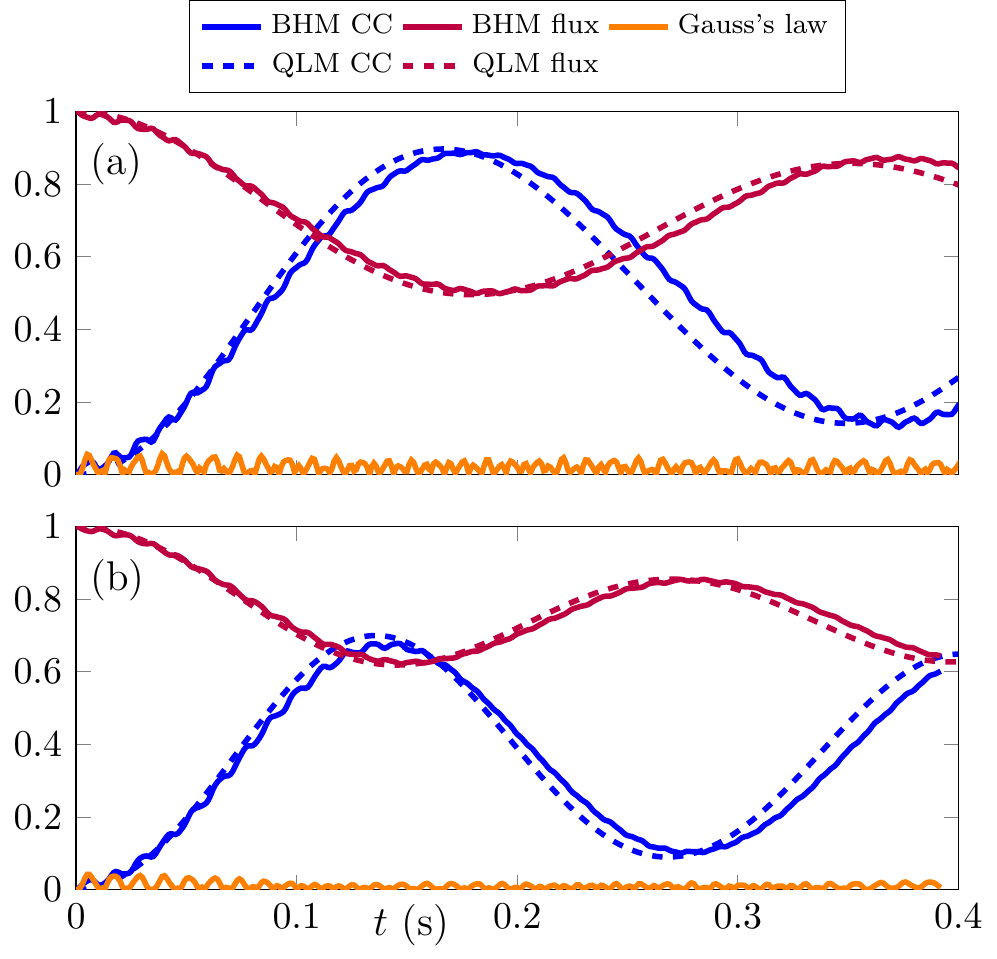}
	\caption{(Color online). Numerical simulations of a global quench of the extreme vacuum state in the BHM~\eqref{eq:eBHM} and target QLM~\eqref{eq:tar}, showing the values of the electric flux~\eqref{eq:flux} and chiral condensate~\eqref{eq:cc}, as well as the gauge violation in the BHM~\eqref{eq:violation}.
    These quenches are in a different parameter regime than Fig.~\ref{fig:quench-ev}, which correspond to a linear architecture where extra renormalization terms~\eqref{eq:inhomog} appear in the Hamiltonian in second-order perturbation theory. Both panels use \(J = 5\)\,Hz, \(\gamma = 57\)\,Hz, \(\delta = -220\)\,Hz, \(V = 160\)\,Hz and \(W = 20\)\,Hz.
    (a) Quench to \(U = 40\)\,Hz, which corresponds to \(\mu/\kappa \approx -0.13\) and \(g^2/\kappa = 0\) in the target model.
    (b) Quench to \(U = 40.5\)\,Hz, which corresponds to \(\mu/\kappa \approx 0.23\) and \(g^2/\kappa \approx -0.96\) in the target model.}
	\label{fig:quench-ev-old}
\end{figure}

\begin{figure}[t!!]
	\centering
	\includegraphics[width=\linewidth]{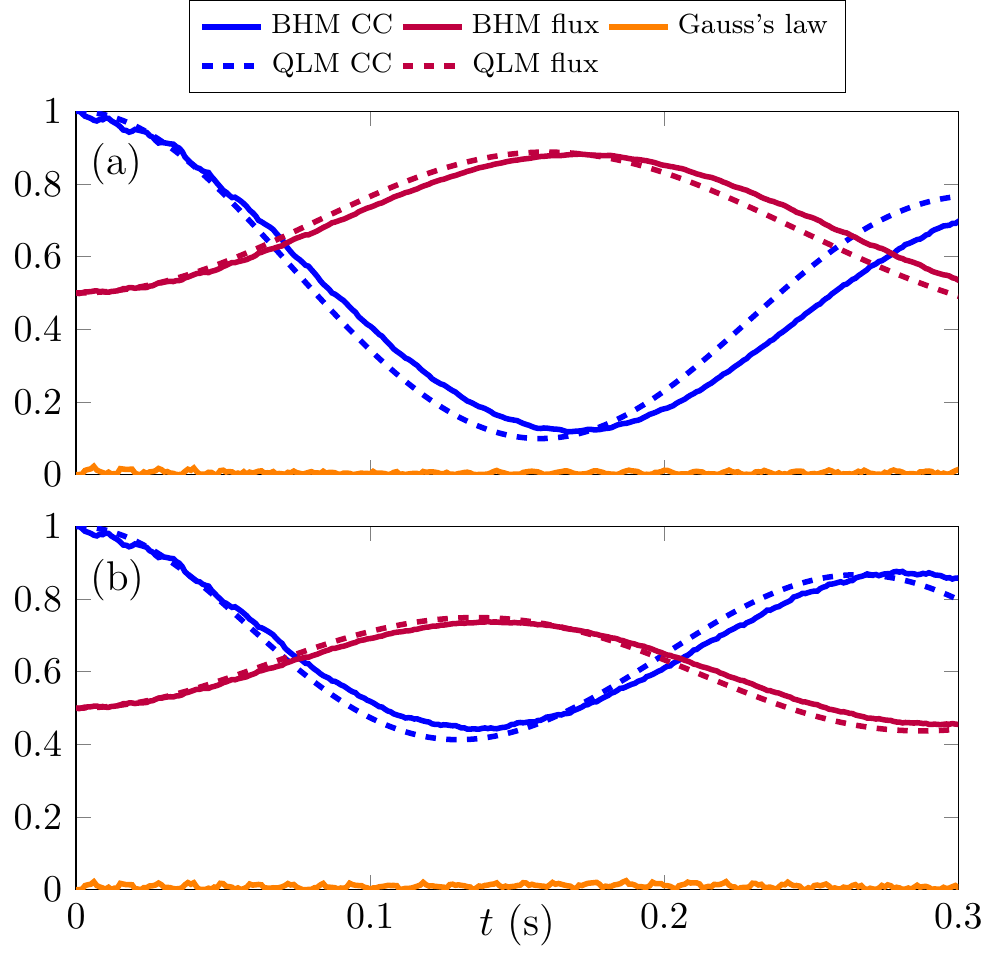}
	\caption{(Color online). Numerical simulations of a global quench of the charge-proliferated state in the BHM~\eqref{eq:eBHM} and target QLM~\eqref{eq:tar}, showing the values of the electric flux~\eqref{eq:flux} and chiral condensate~\eqref{eq:cc}, as well as the gauge violation in the BHM~\eqref{eq:violation}.
    The quench parameters for each panel are the same as in Fig.~\ref{fig:quench-ev-old}.}
	\label{fig:quench-cp-old}
\end{figure}

\begin{figure}[t!!]
	\centering
	\includegraphics[width=\linewidth]{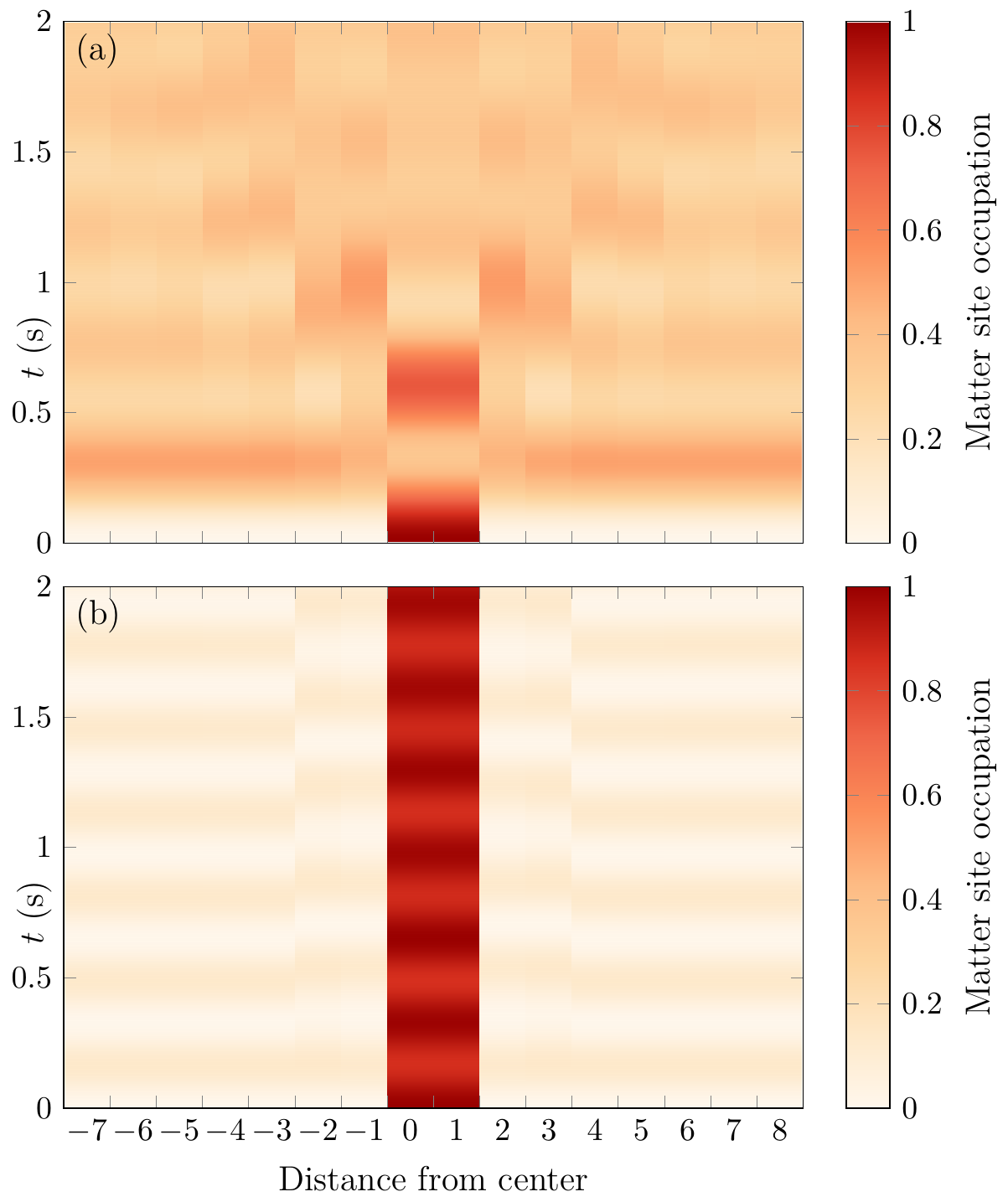}
	\caption{(Color online). Numerical simulations of an electron--positron pair on top of the extreme vacuum in the ideal QLM~\eqref{eq:tar}, described by link operators $\hat{s}^+_{\ell,\ell+1}/\sqrt{2}$ instead of the $\hat{\tau}^+_{\ell,\ell+1}$ used in the main text.
    The same sets of parameters are used as in Fig.~\ref{fig:pair}. The qualitative difference is very small.}
	\label{fig:pair-ideal}
\end{figure}

The four relevant configurations (where the first site corresponds to a gauge site) are \(\ket{21004}\), \(\ket{40012}\), \(\ket{01202}\) and \(\ket{20210}\).
The total energy of each of these configurations is the same as that obtained from the values of \(g^2\) \eqref{eq:g2} and \(\mu\) \eqref{eq:mu} without the renormalization effects from second-order perturbation theory (i.e., the terms proportional to \(J^2\)).
The renormalization for the corresponding terms in the QLM will be different: for the state \(\ket{21004}\) the difference in the renormalized energy from that obtained from the homogeneous configurations is
\begin{equation}\label{eq:21004}
    E^{21004} = J^2 \bigg[ \overbrace{\frac{1}{\delta+2V-6W-\gamma}}^{\ket{21004}\leftrightarrow\ket{20104}} - \overbrace{\frac{1}{\delta+V-4W-\gamma}}^{\ket{21012}\leftrightarrow\ket{20112}} \bigg];
\end{equation}
the difference for \(\ket{40012}\) is the same but with \(-\gamma\) replaced with \(\gamma\).
The difference in the renormalized energy for \(\ket{01202}\) is then (with \(\ket{20210}\)'s difference being obtained similarly by replacing \(\gamma\) with \(-\gamma\))
\begin{align}\label{eq:01202}
    &E^{01202} = J^2 \bigg[ \overbrace{\frac{4}{-\delta+2W+\gamma}}^{\ket{01202}\leftrightarrow\ket{02102}} - \overbrace{\frac{4}{-\delta+V+\gamma}}^{\ket{01210}\leftrightarrow\ket{02110}} \nonumber\\
    &\quad + \overbrace{\frac{3}{-2U+\delta+2V-2W-\gamma}}^{\ket{01202}\leftrightarrow\ket{00302}} - \overbrace{\frac{3}{-2U+\delta+V-\gamma}}^{\ket{01210}\leftrightarrow\ket{00310}} \nonumber\\
    &\quad + \overbrace{\frac{2}{U-\delta-2V+2W-\gamma}}^{\ket{01202}\leftrightarrow\ket{01112}} - \overbrace{\frac{2}{U-\delta-3V+4W-\gamma}}^{\ket{20202}\leftrightarrow\ket{20112}} \bigg].
\end{align}
To account for these energy shifts in the QLM mapping, we add four extra terms to the QLM Hamiltonian which shift the energy for each of these inhomogeneous configurations
\begin{align}
    &H_\text{inhomog} = E^{21004} \sum_{\ell} \frac{1 + \hat{\sigma}^z_\ell}{2} \frac{1 - \hat{\sigma}^z_{\ell+1}}{2} (\hat{s}^z_{\ell,\ell+1})^2 \nonumber\\
    &\quad + E^{40012} \sum_{\ell} \frac{1 - \hat{\sigma}^z_\ell}{2} \frac{1 + \hat{\sigma}^z_{\ell+1}}{2} (\hat{s}^z_{\ell,\ell+1})^2 \nonumber\\
    &\quad + E^{01202} \sum_{\ell} \frac{1 + \hat{\sigma}^z_\ell}{2} \frac{1 - \hat{\sigma}^z_{\ell+1}}{2} \left[1 - (\hat{s}^z_{\ell,\ell+1})^2\right] \nonumber\\
    &\quad + E^{20210} \sum_{\ell} \frac{1 - \hat{\sigma}^z_\ell}{2} \frac{1 + \hat{\sigma}^z_{\ell+1}}{2} \left[1 - (\hat{s}^z_{\ell,\ell+1})^2\right]. \label{eq:inhomog}
\end{align}

Unfortunately, the terms in \(H_\text{inhomog}\) can severely limit the range of dynamics in the QLM.
However, by inspecting the expressions for the energy shifts in Eqs.~\eqref{eq:21004} and \eqref{eq:01202}, we can see that each term comes in a pair with a difference of \(V-2W\) in the denominators.
By setting \(V = 2W\), the two terms in each pair will cancel each other out, removing the second-order energy shifts of these configurations, and thus retrieving a QLM mapping without the restrictive terms in \(H_\text{inhomog}\)~\eqref{eq:inhomog}.
To obtain \(V = 2W\) in an experimental setup, we propose using a zigzag architecture, as discussed in Sec.~\ref{sec:experiment}.

\section{Numerical details}\label{app:MPS}
We use matrix product state (MPS) techniques~\cite{Uli_review,Paeckel_review,mptoolkit} to simulate the quench dynamics of the uniform states and electron--positron pair states.
The time evolved states are obtained using the time-dependent variational principle (TDVP) algorithm~\cite{Haegeman2011,Haegeman2016}: we use a single-site evolution scheme with adaptive bond dimension growth~\cite[App.~B]{vumps}, and use a time-step of \(0.1\)\,ms.
For the simulation of the inhomogeneous states containing an electron--positron pair on top of a vacuum, we used a finite system with a width of at least $40$ matter sites, so that any boundary effects would not be visible in the evolution times considered.
For the simulation of the uniform initial states, we write the state as an infinite MPS with a translation-invariant four-site unit cell, allowing us to simulate the dynamics directly in the thermodynamic limit.

In order to represent the linear tilt term \(\sum_j j\gamma \hat{n}_j\) in the BHM acting on a translation-invariant state, we perform a transformation to the ``dynamic gauge''~\cite{zisling2022}, where the tilt term is removed and replaced by a time-dependent phase in the hopping term \(\sum_j (\mathrm{e}^{\mathrm{i}\gamma t} \hat{b}_j^\dagger \hat{b}_{j+1} + \text{H.c.})\).
This makes the extended BHM amenable to be simulated using infinite MPS techniques, while leaving the expectation values of boson occupation numbers unchanged.

\section{Linear architecture}\label{app:linear}
In the main text, we have focused on the zigzag architecture outlined in Sec.~\ref{sec:experiment}. We can also utilize the linear architecture, also described in Sec.~\ref{sec:experiment}, but then we are restricted to $W=V/2^3$. This leads to the renormalization terms described in Appendix~\ref{app:inhomog}, which modify the form of the target QLM. We shall set $J=5$\,Hz, $\gamma=57$\,Hz, $V=8W=160$\,Hz, and $\delta=-220$\,Hz for this architecture.

We now benchmark the BHM quantum simulator based on the linear architecture by quenching the extreme vacuum, already used in Sec.~\ref{sec:EV} and~\ref{sec:CP}, but with the BHM at $U=40$ ($40.5$)\,Hz, which corresponds to a quench by the target QLM with the renormalization term~\eqref{eq:inhomog} at $\mu/\kappa\approx-0.13$ ($0.23$) and $g^2/\kappa\approx0$ ($-0.96)$. The corresponding quench dynamics of the electric flux and chiral condensate are shown in Fig.~\ref{fig:quench-ev-old}(a,b) for both sets of parameters, respectively. In both quenches, we find very good quantitative agreement in the dynamics of both observables between the BHM and QLM, with a very suppressed and well-controlled gauge violation over all investigated times.

For completeness, we repeat the same quench protocols but starting in the charge-proliferated state considered in Sec.~\ref{sec:CP}, with the corresponding quench dynamics shown in Fig.~\ref{fig:quench-cp-old}(a,b), respectively. The conclusion is the same as that of Fig.~\ref{fig:quench-ev-old} in that the BHM and QLM simulations show very good quantitative agreement in the electric flux and chiral condensate. The gauge violation in the BHM simulation is quite small and well-controlled over all accessible evolution times.

\section{Comparison to ideal QLM}\label{app:idealQLM}
As discussed in the main text, the gauge field $\hat{\tau}^+_{\ell,\ell+1}$ in the target QLM~\eqref{eq:tar} differs from $\hat{s}^+_{\ell,\ell+1}$ of the ideal QLM~\eqref{eq:QLM}. This can lead to quantitative differences in certain regimes, but it is interesting to check the qualitative effect this different representation will have on a relevant high-energy phenomenon like the (de)confinement of the electron--positron pair studied in Fig.~\ref{fig:pair} in the main text. To this effect, we repeat the quenches of the target QLM in Fig.~\ref{fig:pair} but with the ideal QLM Hamiltonian~\eqref{eq:QLM}. The corresponding result, presented in Fig.~\ref{fig:pair-ideal}, shows little qualitative difference in the (de)confinement of the electron--positron pair between quenching with the ideal or target QLM.

\bibliography{biblio}
\end{document}